\documentstyle[aps,multicol,epsfig]{revtex}
  
\begin{document}
\draft

\title{Pattern formation in a predator-prey system characterized by a spatial scale of interaction}

\author{E. Brigatti $^{1,\star}$, M. Oliva
          $^{2}$, M. N\'u\~{n}ez-L\'opez$^{3}$, R. Oliveros-Ramos$^{4}$
         and J. Benavides$^{5}$}  
  
\address{$^{1}$Centro Brasileiro de Pesquisas F\'{\i}sicas, Rua Dr. Xavier 
  Sigaud 150, 22290-180, Rio de Janeiro, RJ, Brasil; \\
  Instituto de F\'{\i}sica, Universidade Federal Fluminense, Campus da Praia Vermelha, 24210-340, Niter\'oi, RJ, Brasil}
\address{$^{2}$ Facultad de F\'{\i}sica, Universidad de La Habana, Ave. Universidad y Ronda, Vedado, 10400,
Havana, Cuba}
\address{$^{3}$ Instituto Mexicano del Petr\'oleo,
Eje Central L\'azaro C\'ardenas Norte 152, Gustavo A. Madero, 07730
DF, M\'exico}
\address{$^{4}$ Centro de Investigaciones en Modelado Oceanogr\'afico y Biol\'ogico Pesquero,  IMARPE, Apartado 22, Callao, Per\'u }
\address{$^{5}$ Institut des Sciences de l'Evolution, Place Eugene Bataillon, CC 065, 34095 Montpellier, France}
\address{$*$e-mail address: edgardo@cbpf.br}

\maketitle
\widetext
  
\begin{abstract}
We describe pattern formation in 
ecological systems using a version of the classical 
Lotka-Volterra model 
characterized by a spatial scale which controls the predator-prey interaction range.
Analytical and simulational results show that patterns can emerge in 
some regions of the parameters space where the instability is driven by 
the range of the interaction. 
The individual-based implementation captures realistic ecological 
features. In fact, spatial structures emerge in an erratic oscillatory regime 
which can contemplate predators' extinction.
\end{abstract}

   
\begin{multicols}{2}

The study of spatial aspects of population dynamics,
with the analysis of patterns and the mechanisms determining 
spatial correlations, started with a seminal work by Moran \cite{moran}.
Afterwards, theoretical 
studies highlighted how different synchronizing mechanisms 
might interact to produce spatial patterns \cite{biospace}.
At present,  a central issue in population ecology is whether the
predator-prey interactions can  
be considered among these mechanisms. 
In fact, spatial correlations between preys and predators can be observed in real populations
and a clear empirical evidence for this phenomenon
can be found, for example,  in a system involving predatory beetles and larval flies as their 
preys\cite{tobin}. 
Generally, these systems have been theoretically 
described using Lolka-Volterra inspired models which are
capable of generating diffusion-driven instabilities. 
The analysis of these reaction-diffusion models argued 
that development of  spatial patterns is possible only under some 
general conditions \cite{murray89} 
regarding the values of the diffusion coefficients
 (predator disperses faster than the prey) and 
regarding the type of the growth functions and predator functional response\cite{holmes}. 

In this paper we are interested in showing
how different and more realistic mechanisms can generate spatial inhomogeneities
which are not directly produced by diffusion phenomenon. 
Standard Lotka-Volterra equations describe a two-species system 
where  both species can coexist  with the population
densities regularly oscillating in time.
This result is similar to empirical observations
 and it is a remarkable prediction considering the simple mathematical elegance which characterizes the model.
With the aim of preserving the lightness 
and economy of this approach, we will introduce a 
simple modification of the Lotka and Volterra's original idea in a spatial version of their model.
The new ingredient is a probability of interaction (predation) 
which becomes a function of the distance between individuals. 
An intuitive justification rests upon considering that the probability that a consumer meets a prey should be  
dependent on the relative distance between them \cite{ecology}.
The introduction of a spatial scale of interaction has been widely applied to model competition describing  species coevolution in community ecology  \cite{ecology}.
More recently it was used in studies of  evolutionary theory which 
propose models for the  emergence of polymorphism or sympatric speciation  \cite{speciation}.

\begin{figure}
\centerline{\psfig{figure=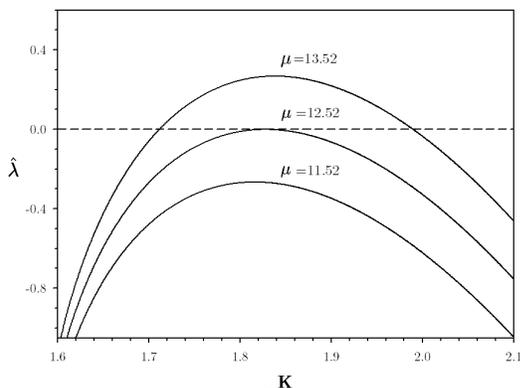, width=6.8cm, angle=0}}
\caption{ \small Dispersion relation $\widehat{\lambda}(K)$ for different
  values of the parameters ($\mu=\sqrt{rm}L_{1}^{2}/D$).}
\label{fig_dis}
\end{figure}

We consider a model characterized by a couple of equations, one for the prey
$N(x,t)$  and one for the predator $P(x,t)$. They describe diffusion in real space 
and the strength of the interaction in the nonlinear term 
is a function of individuals' proximity \cite{lopez}. We do not introduce a single interaction scale $L$
because we consider effective ranges of interaction (the real meeting area can have a different relevance 
for the growth of predators and for the death of preys):
{\scriptsize
\begin{eqnarray}
\frac{\partial N(x,t)}{\partial
t}&=&D_{N}\frac{\partial^{2}N(x,t)}{\partial x^{2}} +rN(x,t)
-\alpha N(x,t) \int_{x-L_{1}}^{x+L_{1}} P(s,t) ds \nonumber \\ 
\frac{\partial P(x,t)}{\partial
t}&=&D_{P}\frac{\partial^{2}P(x,t)}{\partial x^{2}} -mP(x,t)
+\beta P(x,t) \int_{x-L_{2}}^{x+L_{2}} N(s,t) ds.\nonumber \\  
\label{eq_1}
\end{eqnarray} }
Predators consume the preys with an intrinsic rate $\alpha$ and 
reproduce with rate $\beta$;
$r$ is the preys' growth rate and predators are assumed to spontaneously 
die with rate $m$.
$D_N$ and $D_P$ are the diffusion coefficients of preys and predators, respectively.
This system presents  two stationary and spatially homogeneous solutions: an absorbing 
phase $N(x,t)=P(x,t)=0$ and a survival phase
$\bar{N}(x,t)=\frac{m}{2\beta L_{2}}$; $\bar{P}(x,t)=\frac{r}{2\alpha L_{1}}$.
With the aim of investigating the existence of solutions with spatial structure we make a stability 
analysis around $\bar{N}(x,t)$ and $\bar{P}(x,t)$ by considering small harmonic perturbations \cite{murray89,lopez}:
$N(x,t)=\bar{N}+A_{N}exp[\lambda t +i k x]$;
$P(x,t)=\bar{P}+A_{P}exp[\lambda t+i k x]$.
Their introduction into eq.~\ref{eq_1} leads to a linear system 
which exhibits solutions when the determinant equals zero. 
This condition results in the following dispersion relation:
{\scriptsize
\begin{equation}
\lambda(k)
=
-\frac{k^{2}}{2}(D_{N}+D_{P})\pm
\sqrt{\frac{k^{4}}{4}(D_{N}-D_{P})^{2}-rm
\frac{sin(kL_{1})sin(kL_{2})}{k^{2}L_{1}L_{2}}}.
\label{eq:7}
\end{equation}}
This relation shows a symmetry in the interchange of the diffusion constants or the interaction lengths.
Spatial patterns can emerge if the condition $Re[\lambda(k)] > 0$ is satisfied.  
For this reason, $L_{1}\neq L_{2}$ is a necessary prerequisite.
For the case $D_{N}=D_{P}=0$,  the condition obviously allows patterns formation.
This fact proves that the instability is driven by the range of 
the interaction and is independent of the diffusion process. 
It is interesting to note that in this case $N(t)=\sum_{j=1}^n a_{j}(t) \delta(x-j \Delta x_{N})$ 
and  $P(t)=\sum_{j=1}^m b_{j}(t)\delta(x-j \Delta x_{P})$ 
are solutions of the system \cite{perthame}, with $a_{i}(t)$ and $b_{i}(t)$ 
solution of rescaled classical Lotka-Volterra equations.
From these results arise that our instabilities are not diffusion-driven, 
where $D_{N}\neq D_{P}$ is a necessary, albeit not always sufficient condition 
for generating spatial patterns.
It follows that it is natural to simplify our analysis taking $D_{N}=D_{P}=D$.
Moreover, for $L_{2}=2L_{1}$ and  by introducing the rescaled variables 
$K=kL_{1}$ and $\widehat{\lambda}=\lambda\frac{L_{1}^{2}}{D}$, eq.~\ref{eq:7} reduces to:
\begin{equation}
\widehat{\lambda}(K)=-K^{2}+\frac{\sqrt{rm}L_{1}^{2}}{DK}\sqrt{-\sin^{2}K\cos K}.
\label{eq:8} 
\end{equation}
This relation is shown in Fig.~\ref{fig_dis}.
The onset of the instability can be identified by the values of the parameters for which the maximum of the curve
becomes zero ($\widehat{\lambda}(K_{m})=0$).
We can compute it numerically and, for the original variables, we obtain:
\begin{equation}
k_{m}\approx \frac{1.82759}{L_{1}}; \qquad \frac{\sqrt{rm}L_{1}^{2}}{D} \gtrsim12.5232. 
\label{eq_const}
\end{equation}
It is easy to extend this analysis to a two dimensional space. There 
the new relations read: $|k|_{m}\approx 2.19535/L_{1}$ and $\sqrt{rm}L_{1}^{2}/D\gtrsim22.4228$.\\

Now we propose a microscopic discrete 
stochastic formulation of the model which allows us to describe the 
role of demographic fluctuations. 
This implementation introduces two relevant differences. 
The first one is due to the role of intrinsic stochasticity 
which causes internal noise. 
The second one is 
specifically related to the discrete nature of individuals which can generate threshold
effects not present in a continuum description where every small amount of the density 
of population is acceptable, an assumption of continuity which is often unrealistic (atto-fox problem \cite{attofox}). 
Individual-based' lattice models for similar 
Lotka-Volterra system have been studied in details in previous works \cite{lattice,tauber}.
In general, the introduction of stochasticity generates a system considerably richer and perhaps even more realistic. 

For the sake of simplicity our individual based model is implemented in a 
1-dimensional system where the following algorithm was carried out.
Simulations start with an initial population of $P_{0}$
predators and  $N_{0}$
preys, randomly located along a ring (periodic boundary conditions)
of length equal to $1$. The different processes of diffusion, reproduction and death are implemented
sequentially by randomly selecting an individual of each population 
(predator or prey). 
The selected action is repeated for a number
of times equal to the size of the corresponding population. 
When all the processes are carried out, a time step ends and the algorithm restarts.
In detail, we are considering five processes: 1) diffusion, where a predator (prey) is randomly selected and
moves some distance, in a random direction, 
chosen from a Gaussian distribution of standard deviation $\sigma$. 
2) predator reproduction, with rate
$\beta N^{x}_{L_{2}}$, where $N^{x}_{L_{2}}$ is the number of preys which are
at a shorter distance than $L_{2}$ from the predator at position $x$. 
3) predator death, with probability $m$.
4) prey reproduction, with probability $r$.  
5) prey death, with rate $\alpha P^{y}_{L_{1}}$, where $P^{y}_{L_{1}}$ 
is the number of predators which are
at a shorter distance than $L_{1}$ from the prey at position $y$.
All the newborns maintain the same location as the parents.
We evaluate $N^{x}_{L_{2}}$ and $P^{y}_{L_{1}}$ using periodic boundary conditions.
If, in eq.~\ref{eq_1}, we measure time 
in units of the simulation time step, the coefficient 
$D$ is related to the discrete model 
through  $D=\sigma^{2}/2$. Birth and death probabilities  are the 
same in the continuous and in the discrete model.
A rigorous derivation which would show that the continuum field equations approximating 
the discrete model correspond to the eq.~\ref{eq_1}, can be obtained by 
using Fock space techniques \cite{lopez,tauber}. 

\begin{figure}
\centerline{\psfig{figure=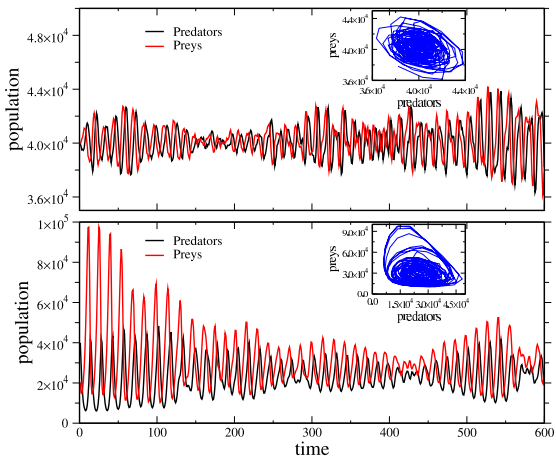, width=8.cm, angle=0}}
\caption{\small Temporal evolution. Top, populations with homogeneous spatial distributions ($L_{1}=L_{2}= 0.1$). Bottom,  populations with modulated spatial distributions ($L_{1}= 0.05$, $L_{2}= 0.1$). Others parameters are: $\sigma=0.004$, $r=m=0.5$, $\alpha=\beta=6.25\times10^{-5}$, $P_{0}=N_{0}=40000$.}
\label{fig_tempo}
\end{figure}

The agent based simulations produced a rich collection of data which allow to
explore the temporal and spacial behavior of our system. 
Irregular oscillations, which swing in a rather erratic 
fashion around an average value, characterize the temporal evolution (see Fig.~\ref{fig_tempo}).
Other stochastic models display a similar behavior \cite{lattice,tauber}.
There, inevitable fluctuations tend to push the system away from 
the trivial survival phase and induce irregular
population oscillations that almost resemble the deterministic cycles of the 
classical Lotka-Volterra model. 
These features well approximate the temporal evolution of our simulations
characterized by homogeneous spatial distributions.
For example, the behavior of the 
dominant Fourier component displayed by the time evolution of these data depend only on the
parameters $r$ and $m$, in a fashion consonant with a classical Lotka-Volterra system.
Moreover, the presence of oscillations is  independent 
of the population size, and therefore they persist in the thermodynamic limit \cite{tauber}.
The more the  spatial solution is marked by clustering
(lower $L_{1}$ and $D$ values), the more the irregularities of the temporal oscillations increase. 
Finally, we must remember that an important outcome of the introduction of intrinsic 
stochasticity is the possibility of predators' extinction. 
For obviously,  when the number of predators becomes very low, a chance fluctuation may lead 
the system into a state with $P(x,t)=0$.
Therefore, asymptotically as $t\to\infty$ this state will be reached.   

\begin{figure}
\centerline{\psfig{figure=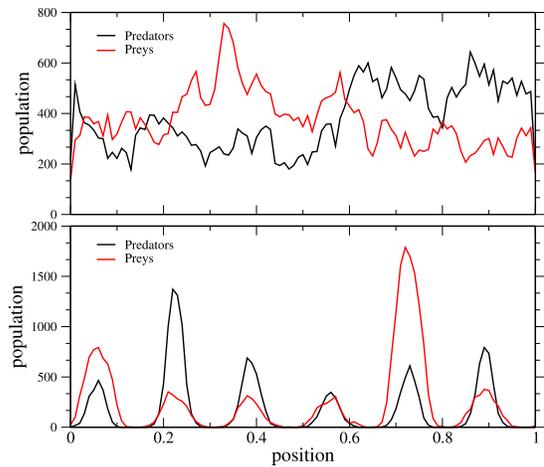, width=8.cm, angle=0}}
\caption{\small Spatial distributions. Top, homogeneous spatial distributions ( $L_{1}=L_{2}= 0.1$).
Bottom, modulated distributions ($L_{1}= 0.05, L_{2}= 0.1$).  We can note preytaxis. 
Same parameters as in Fig.~\ref{fig_tempo}, t=600.
}
\label{fig_mod}
\end{figure}

\begin{figure}
\centerline{\psfig{figure=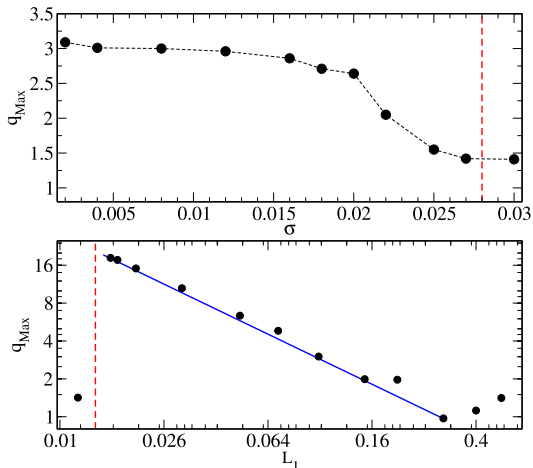, width=8.cm, angle=0}}
\caption{\small Top, $q_{M}$ as a function of $\sigma$ ($L_{1}= 0.1$, $L_{2}= 0.2$, $r=m=0.5$, $P_{0}=N_{0}=20000$). Data are averaged over $50$ realizations. The dashed 
line stands for the analytical result $\sigma_{c}\approx0.028$.
Bottom, log-log plot. 
$q_{M}$ as a function of $L_{1}$ ($L_{2}=2 L_{1}$, $\sigma= 0.004$, $r=m=0.5$, $P_{0}=N_{0}=40000$). 
Data are averaged over $20$ realizations. 
The continuous line represents the analytical prediction: $n\approx0.29/L_{1}$. 
The dashed 
line stands for the analytical result $L_{c}\approx0.014$. The structure functions are averaged over $20$ time steps, starting at $t=580$. }
\label{fig_trans}
\end{figure}

Now we turn to the analysis of the spatial aspects of the distributions
of the populations. 
In accordance with the results of the continuous model 
no regular patterns can emerge 
for interactions characterized by the same range. 
For $L_{1}\ne L_{2}$, clear spatial patterns 
are generated (Fig.~\ref{fig_mod}). 
For an interaction equal to the ring dimension, the simulation is 
characterized by a spatially homogeneous occupancy. 
Decreasing the $L_{1}$ value one peak appears, with 
the population concentrated in one region of the ring. 
For lower values of the interaction length, 
some regions of the ring are occupied forcing all the remaining areas, 
up to some range, to be nearly empty.
This state corresponds to  a sequence
of isolated colonies (spikes) \cite{lopez,edo}.
We also observe that high population density can generate distributions 
where the colonies merge up, without loosing the ordered character of the modulation.
We have deeply explored the case with  $L_{1}=2L_{2}$.
Even if $L_{1}\ne L_{2}$, the number of peaks is generally 
equal for predators and preys and
the tuning of the parameter $L_{1}$ allows
modulations of arbitrary wavelengths. 
Finally, for extremely short-ranged interaction, 
a noisy spatially homogeneous distribution appears (see Fig.~\ref{fig_trans}).
The reported periodic spatial patterns are stationary, with configurations characterized by a 
noise which increases in the neighborhood of the transition towards the homogeneous distribution.
These outcomes obtained from simulations on the segment $(0,1)$ can be 
extended into a 2-dimensional space where fluctuating clusters arranged on
an hexagonal lattice can emerge.

We can observe that for low $D$ values and low population density 
disordered spikes can appear, 
independently on the values of $L_{1}$, probably generated
by the asymmetry between birth and death processes \cite{nature}.
In fact, birth events only occur adjacent to a 
living organism, whereas deaths occur anywhere.  This fact introduces a source
of spatial correlation which can result in a reproductively driven cluster mechanism.
Hence, under reproductive fluctuations and in the presence of a weak
diffusion, individuals can organize into clusters.
These  inhomogeneous configurations can be related with the patterns observed
in discrete lattice simulations, 
where spatial structures can also be generated by traveling waves \cite{tauber,mercedes}.

In the previous paragraph we show how inhomogeneous spatial distributions can appear, depending on the parameters value. 
Now, we will try to characterize the transition towards 
these states (segregation transition).
A proper order parameter is provided by 
$q_{M}=\mathop{max}_{q>0}\Big| \sum \limits_{j=1}^{N(\tau)}\exp[i 2\pi q\cdot x_{j}(\tau)]\Big|^2$,
where the sum is performed over all individuals $j$
with their positions determined by $x_{j}$ at a given time $\tau$.
The transition from a homogeneous to an inhomogeneous distribution matches the 
jump of $q_{M}$ to an integer value, 
corresponding to the number of periodic clusters present in the space. 
In fact, if the space is homogeneously occupied $q_{M} \simeq 1.4$ and
the segregation transition is characterized by the passage of $q_{M}$
from $1.4$ to an integer value as soon as a modulation becomes dominant \cite{edo,lopez}.
In Fig.~\ref{fig_trans} we show $q_{M}$ as a function of $D$ and $L_{1}$. 
For any value of the range of the interaction, a critical value of the diffusion coefficient ($D_{c}$) exists above which no spatial structures emerge.
For $r=m=0.5$ and $L_{1}=0.1$ the analytical prediction gives 
$D_{c}=4\times10^{-4}$, in good accordance with the discrete model 
where the first set of simulations for which $q_{M}=1.4$ presents a diffusion coefficient equal to  $D_{c}\approx3.7\times10^{-4}$. 
Moreover, a critical value of $L_{1}$ exists ($L_{c}$) for which the segregation 
transition takes place. 
For $L_{1}>L_{c}$ 
clusters appear and the value of $L_{c}$ obtained by the simulations
is in accordance with the
analytical result coming from eq.~\ref{eq_const}.
Finally, another correspondence between the predictions of the continuous model and the Monte Carlo simulations exists. In fact, the continuous description can  even reproduce quantitatively the period of the patterns in the modulated distributions. 
For example, considering the case displayed in Fig.~\ref{fig_mod}, for $L_{1}=0.05$, the first relation in~(\ref{eq_const}) tell us that the fastest growing mode is $k=36.6$. 
This is in good agreement with $6\times2\pi=37.7$, which is the wavenumber of  the  
configuration generated by our simulation. 
In fact, this wavenumber is the first immediately above the fastest growing
mode of the mean-field description which is compatible with the periodic
boundary conditions. 
For obvious,  from the same equation in (\ref{eq_const}),
we can obtain the general analytical relation for the number of peaks $n$: $n\approx0.29L_{1}^{-1}$, 
which is compared with the Monte Carlo data in Fig.~\ref{fig_trans}.\\

In conclusion, the introduction of a finite-range interaction in a spatial Lotka-Volterra model, 
allows the description of a rich spatio-temporal dynamics characterized by regular spatial structures.
This type of spatial behavior is  a central issue in population ecology as, effectively,
it is possible to record spatial correlations between preys and predators in nature. 
Our investigation was carried out both analytically, using a suitable continuous model (mean-field
approach), as well with a discrete model, by employing Monte Carlo simulations. 
The individual-based model shows the existence of spatial structures in an erratic oscillatory regime 
which can contemplate predators' extinction.
The mean-field description  captures 
the essential features of the discrete model. We record quantitative correspondences for 
the period of the spatial patterns in the modulated distributions, the  value of the critical diffusion and the critical interaction length. 
From an ecological perspective, the emergence of spatial patterns in an erratic oscillatory regime are realistic elements generally absent from 
conventional approaches and might shed further light on issues of particular ecological relevance. \\

We thank the Brazilian agencies CNPq for partial financial support and the school CSSS 2008, 
of the Santa F\'e Institute, where this work was conceived. 
We are grateful to B. Perthame for helpful discussions.

\end{multicols}

\end{document}